\documentclass[%
reprint,
superscriptaddress,
showpacs,preprintnumbers,
showkeys,
 amsmath,amssymb,
 aps,
]{revtex4-1}
\usepackage{subfigure}
\usepackage{graphicx}
\usepackage{dcolumn}
\usepackage{bm}

\def\be{\begin{equation}}
\def\ee{\end{equation}}
\def\bea{\begin{eqnarray}}
\def\eea{\end{eqnarray}}
\newcommand{\ket}[1]{\mbox{$|#1\rangle$}}
\newcommand{\bra}[1]{\mbox{$\langle#1|$}}

\providecommand{\U}[1]{\protect\rule{.1in}{.1in}}
\begin{document}

\title {Electrically tunable quantum interfaces between photons and spin qubits in carbon nanotube quantum dots}

\author{Ze-Song Shen}
\affiliation{Department of Physics, Center for Optoelectronics Materials and Devices, Zhejiang Sci-Tech University,  Hangzhou, Zhejiang 310018, China}
\author{Fang-Yu Hong}
\affiliation{Department of Physics, Center for Optoelectronics Materials and Devices, Zhejiang Sci-Tech University,  Hangzhou, Zhejiang 310018, China}

\date{\today}
\begin{abstract}
We present a new scheme for quantum interfaces to accomplish the interconversion of photonic qubits and spin qubits based on optomechanical resonators and the spin-orbit-induced interactions in suspended carbon nanotube quantum dots. This interface implements quantum spin transducers and further enables electrical manipulation of local electron spin qubits, which lays the foundation for all-electrical control of state transfer protocols between two distant quantum nodes in a quantum network. We numerically evaluate the state transfer processes and proceed to estimate the effect of each coupling strength on the operation fidelities.
\end{abstract}


\maketitle
\section{INTRODUCTION}
In quantum information processing, quantum interfaces (QIs) play a fundamental role in distributing quantum states between individual nodes of large quantum networks, which also offers potential applications for scalable and distributed quantum computation \cite{hjkb, jcae}. Some promising setups related to quantum networks have been introduced to fulfill information distribution protocols either ``on-chip'' or over long distances \cite{mamp, kspr, ahgs}. They serve to accomplish transferring, swapping, and entangling qubits within the building blocks. The prototype quantum interface (QI) coherently interconverting so-called ``flying'' optical qubits and ``stationary'' matter qubits was proposed by Cirac \textit{et al.} \cite{jcpz}, using cavity-assisted Raman processes. Qubits are stored in hyperfine states of atoms and then transformed into the number states of a photon wavepacket and vice versa \cite{wyrl}. \\
In view of substantial coherence and control of nanoengineered solid-state qubits, recently Stannigel \textit{et al.} \cite{kspr} presented an optical interface with optomechanical resonators \cite{evsd} to address the incompatibility between light and qubits. The scheme makes use of a magnet-mediated spin qubit \cite{prsk} and a charge qubit that employs electrostatic interactions. Analogous magnetic cantilevers to which the spin qubit couples could be utilized for coherent sensing of mechanical resonators \cite{skaj, sdsk}. In the two interfaces above, laser pulses are used as the controlling tool, which requires bulky setups and makes them somewhat inconsistent in large-scale quantum information processing. Here we propose an electrically tunable interface based on a suspended carbon nanotube quantum dot (CNTQD) for implementation.\\
CNTQDs are currently being investigated in a variety of systems, such as vibrational state control \cite{pswb}, ciruit QED \cite{actk}, and coupling to cavity photons \cite{jvmd, mdlb}. The potential for carbon-based systems to be used in quantum information processing has been widely explored by exploiting different degrees of freedom in CNTs and CNTQDs as quantum bits. For example, quantum information can be stored in the motional degrees of freedom of nanomechanical devices \cite{srmh} by creating single quantum excitations (phonons) in a resonator \cite{acmh}. It is also reliable to utilize the spin-orbit interaction with the valley degrees of freedom in CNTQDs \cite{dbbt,kfcm,elfp}.\\
We describe in this paper a quantum interface to realize the all-electrical control of coherent interconversion between optical qubits and the single electron spins in CNTQDs. In our work, we take advantage of the spin degrees of freedom. It is feasible to modify the coupling strength between the electron spin and the vibrational motion of a suspended CNT by shifting the effective center of the quantum dot using a longitudinal electric field \cite{hong}. Our setup proves to be robust and well-controlled owing to the intrinsic spin-mechanical coupling in CNTQD, which could otherwise be used to mechanically manipulate the electron spin \cite{hwgb} supported by the inherent strong spin-orbit interaction \cite{appr}.\\
In addittion, the outstanding mechanical properties of CNT -low mass and noise \cite{irrt}, high quality factor, and widely tunable resonance frequencies \cite{vsyy, ahgs}, make it attractive to be used as a resonator in a QI. Meanwhile, long spin life time of electrons in CNTQDs can be abtained off-resonance \cite{actk}. In the resonant case though, the setup can accomplish qubit control and tranfer while gaining high operation fidelities, with the reduced charge noise-induced dephasing from the suspended CNT \cite{actk} and further elemination of the hyperfine contributions to qubit decoherence.

\begin{figure}

    \centering

    \subfigure
    {
        \includegraphics[width=8cm]{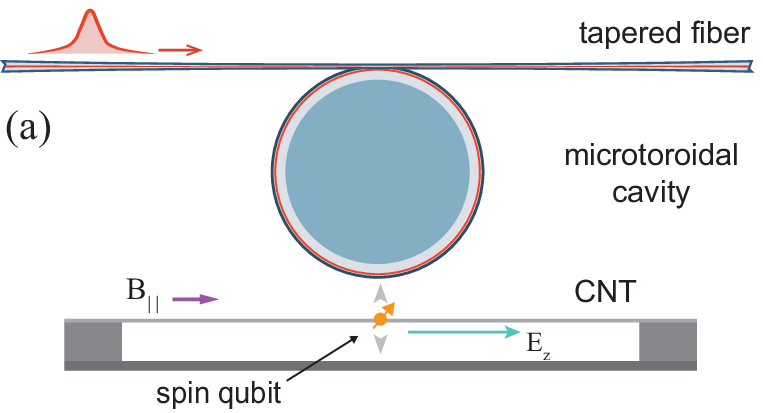}
        \label{fig1a}
    }
    \\
    \subfigure
    {
        \includegraphics[width=8cm]{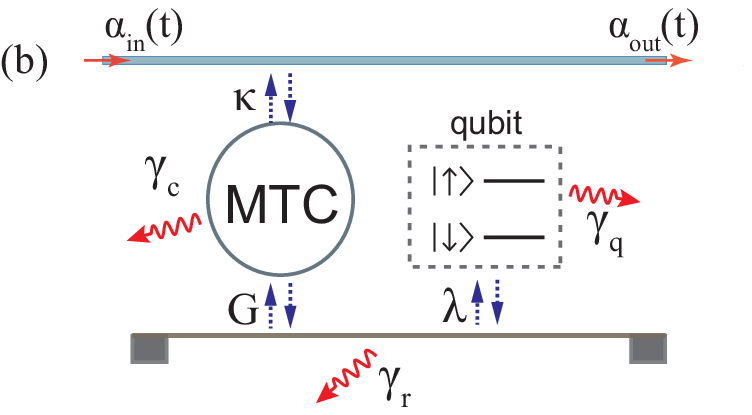}
        \label{fig1b}
    }
\caption{(color online) (a) Schematic quantum interface to achieve robust qubit-light interversion using the spin degree of freedom in a CNTQD: An electron spin ($\uparrow, \downarrow$) is extrinsically coupled to the evanescent optical field in a microtoroidal cavity (MTC) through the intrisinc spin-mechanical coupling, and flying photon quibts are coupled in and out the cavity through a tapered fiber. (b) Illustration of the interactions and the decoherence sources in the system. See text for details.}
    \label{fig1}
\end{figure}

\section{Setup}
To describe the mechanisim of electrical control of our system, we first model the spin of an electron localized in a suspended CNTQD, the relevant mechanical oscillator mode, and the coupling between these two degrees of freedom. \\
In the presence of a longitudinal magnetic field of magnitude around $B^\ast=\frac{\Delta_{so}}{g_s\mu_B}$, the QD can serve as a spin qubit coupled to the mechancial resonator mode with the Hamiltonian \cite{appr}
\be
H_{so}=\frac{\omega_q\sigma_z}{2}+\lambda\left(b^\dagger+b\right)\left(\sigma_+ +\sigma_-\right)+\omega_p b^\dagger b,
\ee
where $\Delta_{so}$ denotes the spin-orbit coupling, $g_s$ is the $g$ factor of the electron spin, $\sigma_z$ is the Pauli operator, $\sigma_{\pm}$ are the raising and lowering operators for the spin qubit with a tunable splitting $\omega_q$, and $\omega_p$ is the frequency of the flexual mode of the CNT described by the annihilation operator $b$. For simplicity, we assume $\hbar=1$ throughout this paper. The spin-phonon coupling $\lambda$ in Fig. \ref{fig1b} is defined as $\lambda=\Delta_{so}\langle f^{'}\rangle \mu_0/2\sqrt{2}$, with $\langle f^{'}\rangle$ being the derivative of the phonon waveform averaged against the electron density profile $n$ in the QD and $\mu_0$ being the zero-point fluctuation amplitude of the phonon mode \cite{appr}. We classify the electron density profile as $n(z)\simeq exp[-A\left(z-z_c\right)^2]$ by assuming a parabolic trapping potential and the ground state of the electron in the CNTQD \cite{asmr}. The parameter $A$ is associated with the effective electron mass $m^\ast$ and the characteristic frequency $\omega_0$ of the harmonic oscillator, and $z_c$ is the center of the potential well. \\
To realize the manipulation of the coupling strength $\lambda$ electrically, we can modulate the electron distribution in the QD by adding and tuning an electric field $E_z(t)$ along the $z$ direction of CNT without modifying the energy level spacing of the QD \cite{hong, jdwa}.\\
Now we proceed to discuss the framework with the union of the spin-oscillator system we described above by considering a single node of an optical network as illustrated in Fig. \ref{fig1a}. 

The total Hamiltonian is given by \cite{ga, smdv}
\bea
H&=&H_0+H_I, \\
H_0&=&\frac{1}{2}\omega_q\sigma_z+\omega_pb^\dagger b+\Delta_cc^\dagger c+\int_0^\infty \omega a_\omega^\dagger a_\omega d\omega, \\
H_I&=&\left(\frac{\lambda}{2}\sigma_+ b+\text{H.c.}\right)+\left(Gc^\dagger+G^\ast c\right)\left(b^\dagger+b\right)\\
&+&\int_0^\infty d\omega\left(\kappa ca_\omega^\dagger +\text{H.c.}\right),
\eea
where $a_\omega$ is the destruction operator for the mode of frequency $\omega$ in the optical quantum channel, and $c$ is the annihilation operator for the cavity mode, respectively. The detuning $\Delta_c=\omega_c-\omega_L-2|G|^2/\omega_p$ and the coupling G can be regulated by the field amplitude and the frequency $\omega_L$ of local driving lasers. The cavity mode is assumed to be coupled to the field in the tapered fiber with a constant $\kappa=\sqrt{\gamma/2\pi}$ in Fig. \ref{fig1b}.\\
Corresponding to the rotating-wave approximation (RWA), we simplify the interaction Hamiltonian by dropping the energy unconserving terms and abtain
\be
\tilde{H}_I=\frac{\lambda}{2}\sigma_+ b+Gc^\dagger b+\int_0^\infty d\omega\sqrt{\frac{\gamma}{2\pi}} ca_\omega^\dagger+\text{H.c.}.
\ee
Under the condition $k_BT \ll \omega_p$, the system has two invariant Hilbert subspaces  with the bases $\{\ket{\downarrow,0,0}\ket{vac}\}$ and $\{\ket{\uparrow,0,0}\ket{vac},\ket{\downarrow,1,0}\ket{vac},\ket{\downarrow,0,1}\ket{vac}, \ket{\downarrow,0,0}a_\omega^\dagger \ket{vac}\}$, respectively. Here in $\ket{k,l}$, $k$ and $l$ denote the number of phonons in the flexual mode and the number of  photons in the cavity, respectively. $\ket{vac}$  denotes the vacuum state of the optical continum. \\
In the interaction picture, the general state of the system could be expressed by the superposition 
\be
\ket{\Psi}=C_0\ket{\downarrow,0,0}\ket{vac}+C_1\ket{\Psi_I}, 
\ee
where
\be
\ket{\Psi_I}=\sum_nc_n\ket{\mu_n}e^{-i\omega_nt},
\ee
and $c_n$ denotes the amplitudes $\alpha_\omega$ and $\beta_{q,r,c}$ of the four relevant states $\ket{\mu_n}$ in the excited Hilbert subspace. In this quantum interface, we have the evolution equations
\be
\dot{c}_n=-i\sum_m\bra{n}\tilde{H}_I\ket{m}e^{i\omega_{nm}t}c_m.
\ee

By expressing the qubit-mechanical coupling in the state amplitudes of the QI system, we abtain the target control pulse
 \be
\lambda=-\frac{2}{\beta_q}\left(\dot{\beta}_r+G\beta_c\right),
\ee
with $\beta_q(t), \beta_r(t), \beta_c(t)$ set according to the specified incoming/outgoing photon wavepacket.
For simiplicity, in this work we tune this ideal control pulse to drive the real system with certain decoherences. Note that careful designs of the manipulation could help to achieve a higher fidelity of quantum transmission. 
\section{Quantum state tranfer}
The mapping process from the stationary spin quibt to the flying photon qubit requires the initial conditions $\alpha_{in}(t_i)=0$, $\beta_q(t_i)=1$, $\beta_r(t_i)=0$, $\beta_c(t_i)=0$. Following the driven evolution passage: $\beta_q\xrightarrow{\lambda_s(t)}\beta_r\xrightarrow{G}\beta_c\xrightarrow{\kappa}\alpha_\omega$, the coherent tranfer of the quantum amplitude is finished, and vice versa. With the normalization understood, the outgoing photon wavepacket $\tilde{\alpha}_{out}(t)$ can be arbitrarily specified to contain an average number of $\sin^2 \theta$ : $\sin^2\theta\int_{t_i}^{t_f}|\tilde{\alpha}_{out}(t)|^2dt=\sin^2 \theta$. At the remote future $t_f\rightarrow +\infty$, the photon generation is completed, \textit{i.e.}, $\beta_r(t_f)=\beta_c(t_f)=0$, hence $\beta_q(t_f)=e^{i\phi}\cos\theta$. 

The general photon-generation process in the QI reads
\bea
&&\left(C_0\ket{\downarrow}+C_1\ket{\uparrow}\right) \ket{vac}\xrightarrow{\lambda_s(t)}C_0\ket{\downarrow}\ket{vac}\\
&&+C_1\left(e^{i\phi}\cos\theta\ket{\uparrow}\ket{vac}+\sin\theta\ket{\downarrow}\ket{\tilde{\alpha}_{out}}\right).
\eea
When the emission operation is completed, $\theta=\pi/2$ and $\alpha_\omega(t_f)=0$. Then, the spin qubit is mapped onto the flying qubit by the conversion
\be
\left(C_0\ket{\downarrow}+C_1\ket{\uparrow}\right) \ket{vac}\xrightarrow{\lambda_s(t)}\ket{\downarrow} \left(C_0\ket{vac}+C_1\ket{\tilde{\alpha}_{out}}\right).
\ee
The absorption of an incoming photon at an receiving node is typically the reverse of the above.\\
Also, the sending manipulation can be mediated such that $\theta<\pi/2$. The initia state $\ket{\uparrow}\ket{vac}$ is then transformed into an entangled state
\be
\ket{\uparrow}\ket{vac}\xrightarrow{\lambda^{'}_s(t)}e^{i\phi}\cos\theta\ket{\uparrow}\ket{vac}+\sin\theta\ket{\downarrow}\ket{\tilde{\alpha}_{out}}.
\ee

If we manage to absorb the above entangled photon, we would have an entanglement between the sending node 1 and the receiving node 2:
\bea
{\ket{\uparrow}}_1{\ket{\downarrow}}_2\xrightarrow[\lambda_r(t)]{\lambda^{'}_s(t)}e^{i\phi}\cos\theta{\ket{\uparrow}}_1{\ket{\downarrow}}_2+\sin\theta{\ket{\downarrow}}_1{\ket{\uparrow}}_2.
\eea

\begin{figure}[t]
\includegraphics[width=8cm]{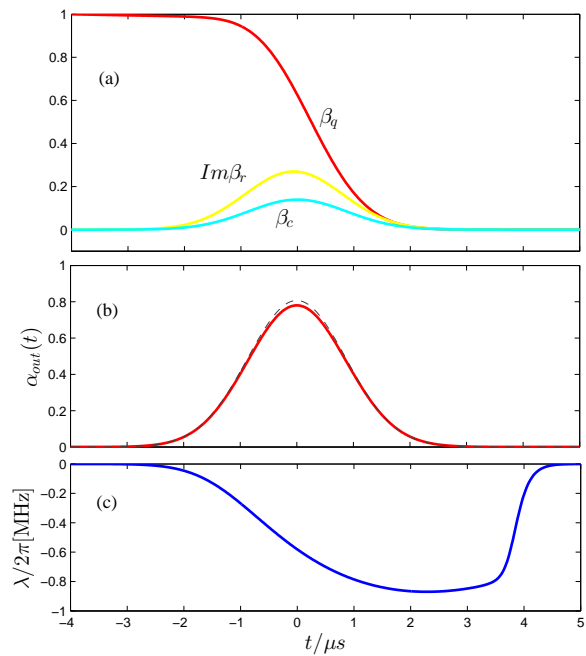}
\caption{\label{fig2}(color online)Generation of a single photon with the shape of a Gaussian wavepacket $\tilde{\alpha}(t)=\text{exp}(-\Gamma t^2)$. (a) The time evolution of state amplitude $\beta_q(t)$, $\beta_c(t)$ and the imaginary $\beta_r(t)$ in the QI. (b) The generated photon wavepacket (solid line) in comparision with the target one (dashed line). (c) The coupling strength $\lambda(t)$ determinining the controlling gate potential for the sending node. The parameters are $\Gamma=0.15\pi\times\sqrt{2}$MHz, $\gamma/2\pi=5$MHz,  $G/2\pi=1.3$MHz, $\gamma_r/2\pi=2.6$KHz, $\gamma_c/2\pi=2.5$KHz and $\gamma_q=0.01$MHz.}
\end{figure}

\section{Numerical Simulation}
Here we evaluate the performance of the QI by taking some necessary decoherences into account. The phonon mode operator obeys the following quantum Langevin equation
\be
\frac{d}{dt}b=-i\left[b,H\right]-\frac{\gamma_r}{2}b-\sqrt{\gamma_r}\zeta(t),
\ee
where $\gamma_r$ denotes the mechanical damping rate and the noise operator $\zeta$ satisfies $\left<\zeta(t)\right>=0$ in thermal equilibrium. At low temperatures $k_BT \ll \omega_p$, we have
\be
\frac{d}{dt}\left<b^{\dagger}b\right>=-i\left<\left[b^{\dagger}b,H-i\frac{\gamma_r}{2}b^{\dagger}b\right]\right>.
\ee
Under the same assumption, the heating of the nanotube resonator from the thermal bath reservoir is negligible and we find the effective interaction Hamiltonian to be ${\tilde{H}_I}^{eff}=\tilde{H}_I-i\frac{\gamma_r}{2}b^{\dagger}b$. Meanwhile, the mechanical damping rate can be derived from $\gamma_r=\frac{\omega_p}{Q_m}$ with $Q_m$ being the quality factor of the mechanical resonator. Here, we take the resonance frequency $\omega_p/2\pi$ as 360MHz and the quality factor $Q_m=140,000$ with the resulting decay rate $\gamma_r/2\pi=2.6$kHz \cite{ahgs} in our simulations. With high quality factors of both cavity and CNT achievable today \cite{hjkb, bmly, kjv, ahgs, gsah}, we assume both the mechanical decay rate $\gamma_r$ and the cavity leakage $\gamma_c$ are small. The long spin life time in the QD is also expected off-resonance due to the low density of states at the spin energy splitting for the phonon spectrum of a suspended CNT. Therefore, taking into account the intrinsic spin qubit dephasing $\gamma_q$ and the cavity decoherence, the system dynamics can then be described by the effective non-unitary Hamiltonian
\bea
\tilde{H}^{eff}&=&H_0+{\tilde{H}_I}^{eff},\\
{\tilde{H}_I}^{eff}&=&\tilde{H}_I-i\frac{\gamma_r}{2}b^{\dagger}b-i\frac{\gamma_c}{2}c^{\dagger}c-i\frac{\gamma_q}{2}\ket{\uparrow}\bra{\uparrow}.
\eea

Correspondingly the evolution equations of the QI are expressed in the following complete form:
\bea
\dot{\beta}_q&=&-i\frac{\lambda}{2}\beta_r-\frac{\gamma_q}{2}\beta_q, \\
\dot{\beta}_r&=&-i\left(\frac{\lambda^\ast}{2}\beta_q+G^\ast\beta_c\right)-\frac{\gamma_r}{2}\beta_r, \\
\dot{\beta}_c&=&-iG\beta_r-\sqrt{\gamma}\alpha_{in}(t)-\frac{\gamma}{2}\beta_c-\frac{\gamma_c}{2}\beta_c, \\
&=&-iG\beta_r-\sqrt{\gamma}\alpha_{out}(t)+\frac{\gamma}{2}\beta_c-\frac{\gamma_c}{2}\beta_c, 
\eea
where we have assumed the resonance condition $\Delta_c=\omega_p=\omega_q$ for the convenience of simulation. Next we use a simple Gaussian wavepacket $\tilde{\alpha}(t)=\text{exp}(-\Gamma t^2)$ with normalization understood, to simulate numerically the functions of the QI.\\

\begin{figure}
    \centering
    \subfigure
    {
        \includegraphics[width=8cm]{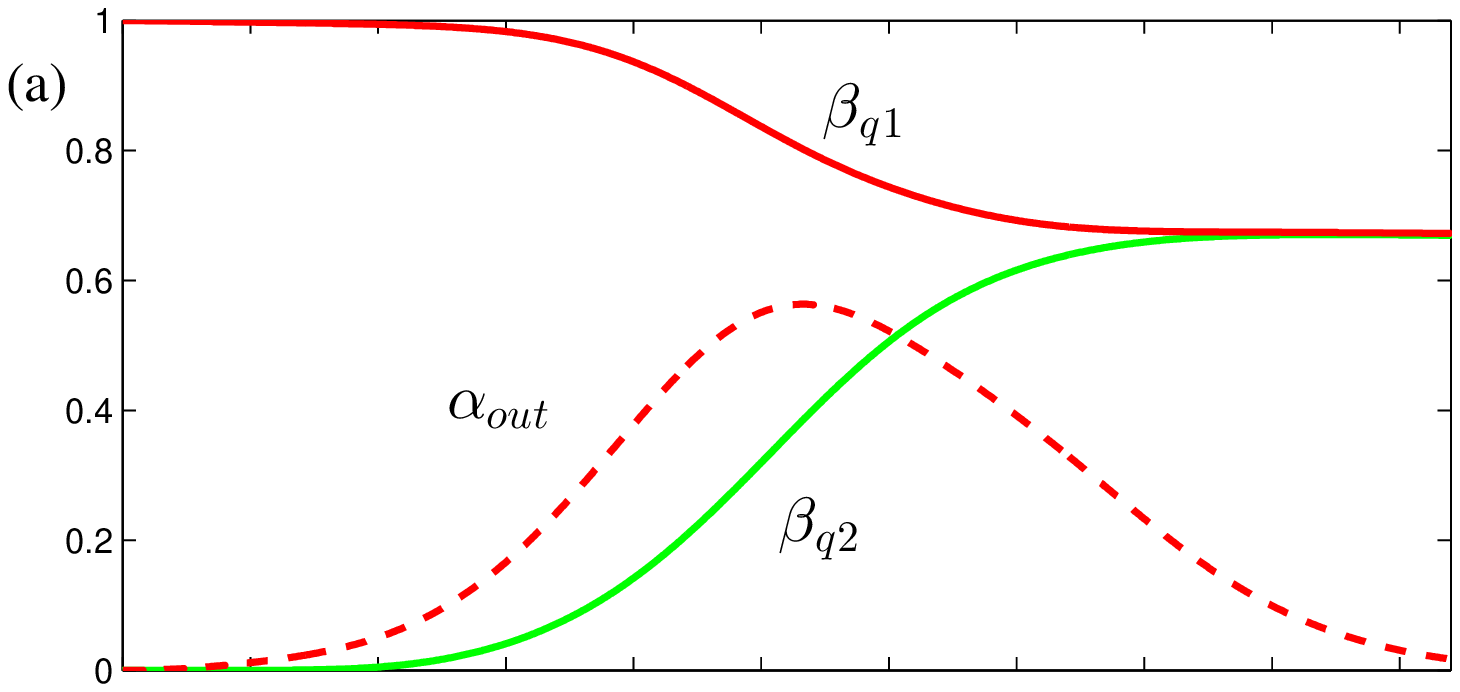}
        \label{fig3a}
    }
  
    \subfigure
    {
        \includegraphics[width=8cm]{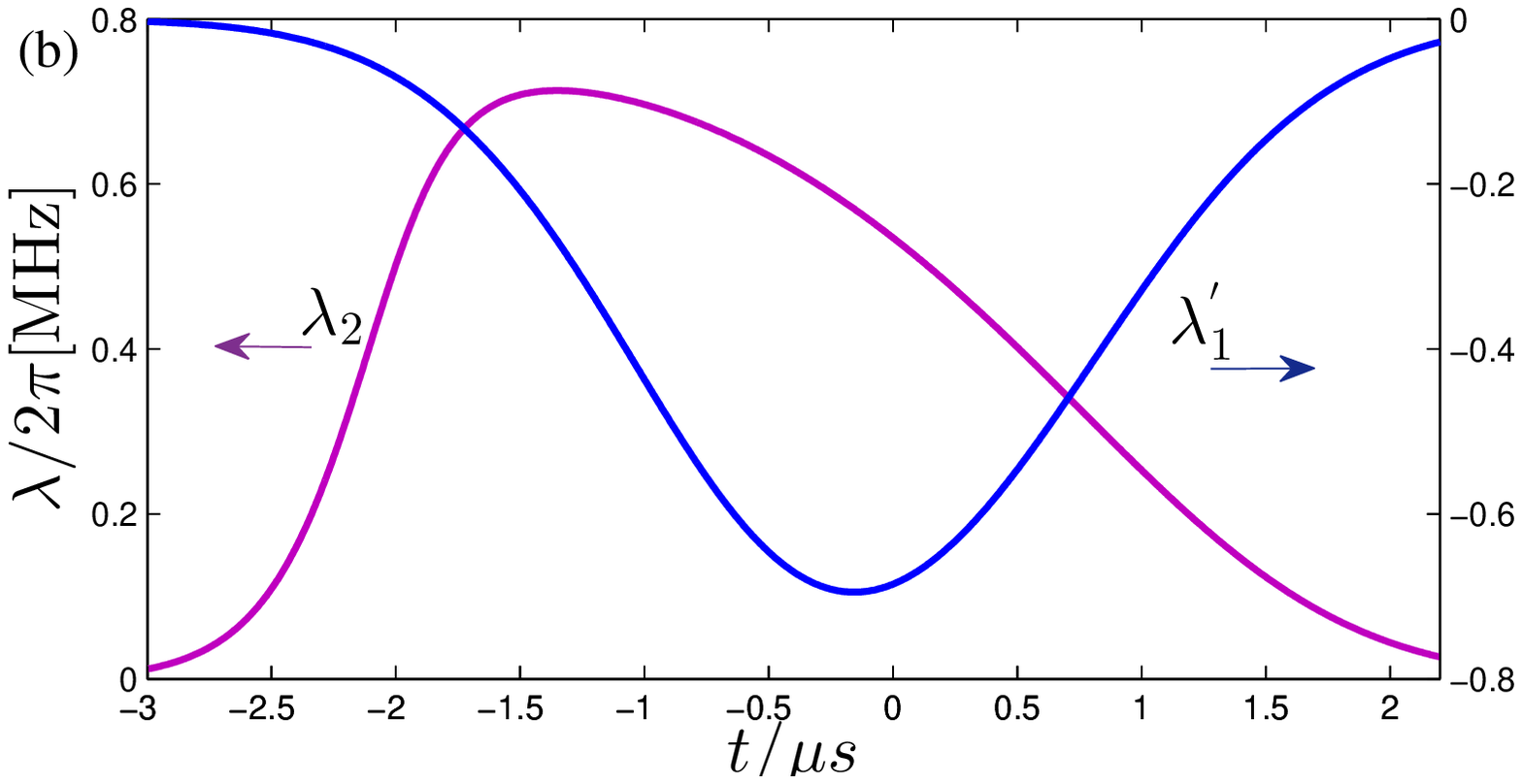}
        \label{fig3b}
    }
\caption{(color online) Distribution of the target entanglement $ \frac{1}{\sqrt2}{\ket{\uparrow}}_1{\ket{\downarrow}}_2+\frac{1}{\sqrt2}{\ket{\downarrow}}_1{\ket{\uparrow}}_2$ between the sending node 1 and the receiving node 2. (a) The evolution of state amplitude $\beta_{q1}(t)$ in node 1 and $\beta_{q2}(t)$ in node 2, both are mediated by the intermediate photon wavepacket $\alpha_{out}(t)$. (b) The driving frequency $\lambda^{'}_1(t)$ for the sending node and $\lambda_2(t)$ for the receiving one. The parameters are $\gamma_1/2\pi=0.5$MHz, $G_1/2\pi=0.9$MHz for node 1,  $\gamma_2/2\pi=4.3$MHz, $G_2/2\pi=1.1$MHz for node 2, and $\gamma_r/2\pi=2.6$KHz, $\gamma_c/2\pi=2.5$KHz, $\gamma_q=0.01$MHz for both.}
    \label{fig3}
\end{figure}

The generation process of a single photon is illustrated in Fig. \ref{fig2}. During this operation, the control strength $-\lambda$ needs to be smaller than the maximum one tunable by mediating the longitudinal electric field strength \cite{hong} to perform this manipulation. 

\begin{figure*}
    \centering
    \subfigure
    {
        \includegraphics[width=5.5cm]{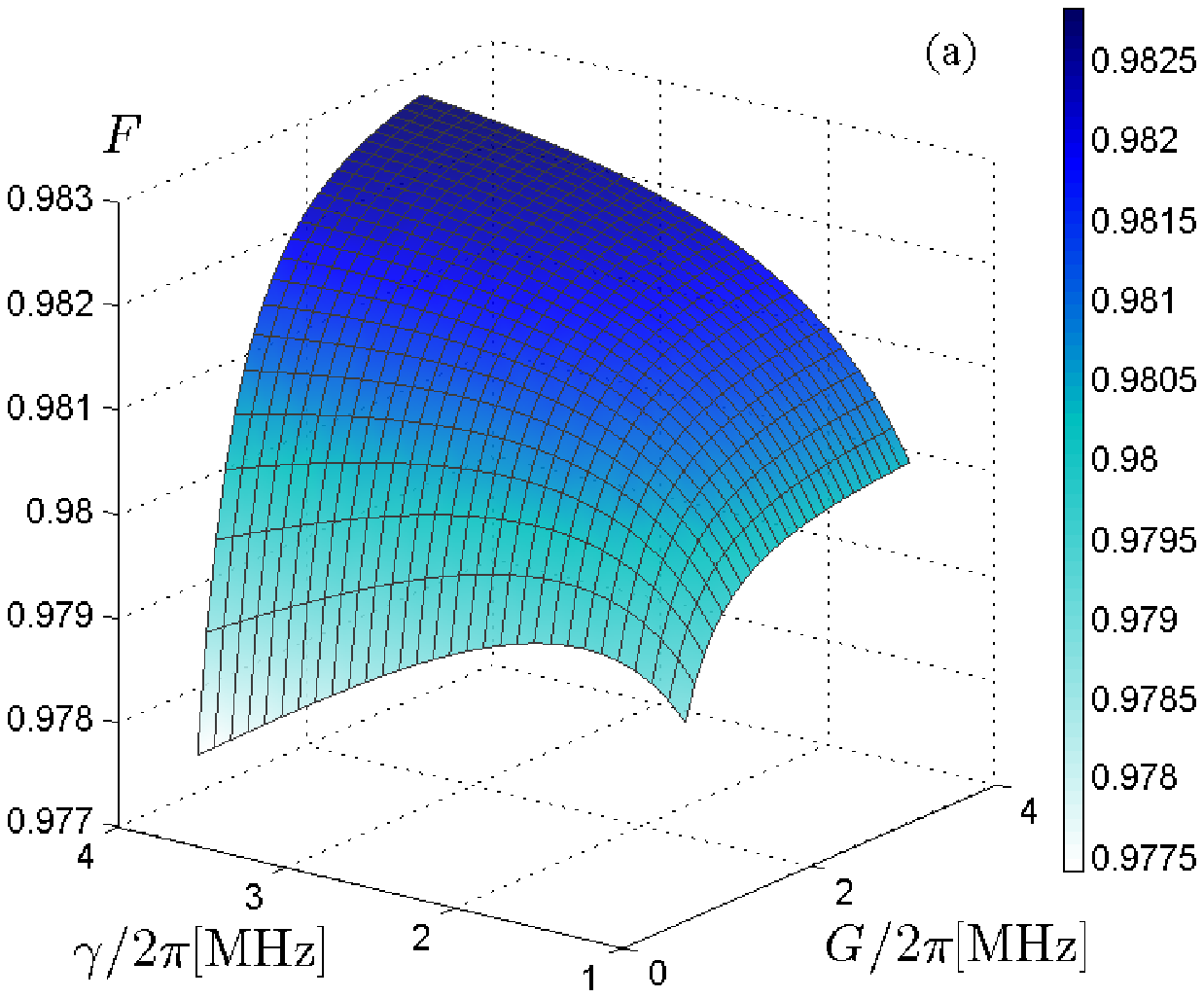}
        \label{fig4a}
    }
    \subfigure
    {
        \includegraphics[width=5.5cm]{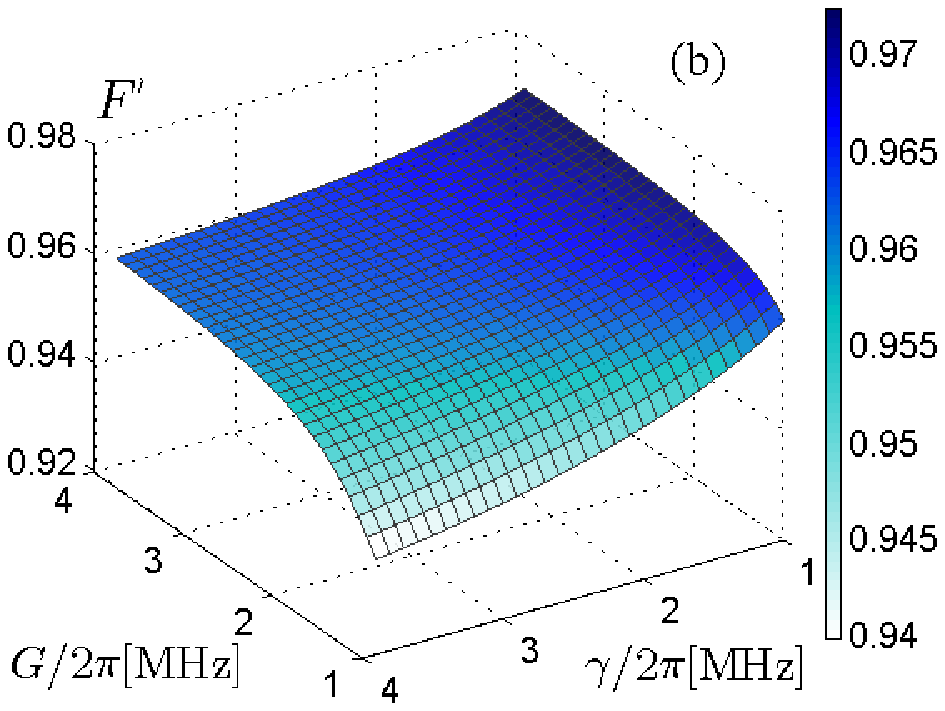}
        \label{fig4b}
    }
    \subfigure
    {
        \includegraphics[width=5cm]{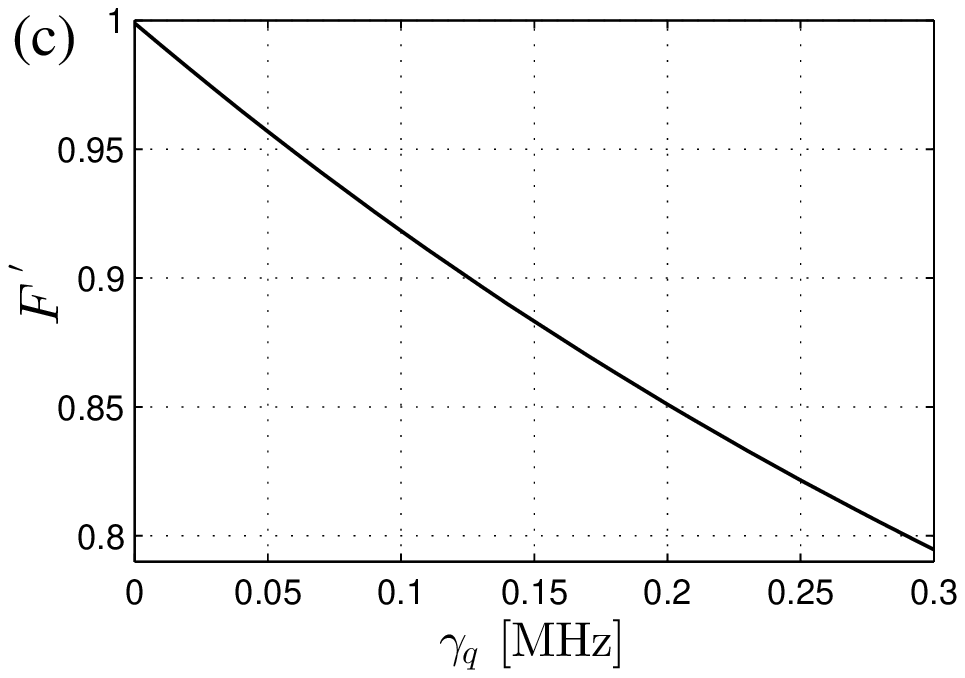}
        \label{fig4c}
   }
\caption{(color online) (a) The relation of  coupling strength $\gamma$, $G$ with the fidelity of generating a Gaussian-shape single photon. The parameters are $\gamma_q=0.01$MHz, $\gamma_r/2\pi=\gamma_c/2\pi=0.01$MHz. (b) The influence of $\gamma$ and $G$ on the fidelity of transferring the target state: $\frac{1}{\sqrt2}\ket{\uparrow}\ket{vac}+\frac{1}{\sqrt2}\ket{\downarrow}\ket{\tilde{\alpha}_{out}}$, from one node to the other. (c) The effect of decoherence source $\gamma_q$  on the fidelity of transferring the above entangled state.  The parameters used in this case are $\gamma/2\pi=G/2\pi=1$MHz, $\gamma_r=\gamma_c=0$.}
    \label{fig4}
\end{figure*}

The parameters used here are within the reach of present state-of-the-art techniques \cite{kspra, appr}. Significantly, the pulse duration is rather short, about $6\mu s$, due to the strong coupling $\gamma$. \\
Note also that meticulous designs of the coupling strength $\gamma$, $G$ and $\lambda$ can help to achieve faster sending and receiving manipulations while ensuring high fidelities. The related estimation is shown in Fig. \ref{fig4}. From Fig. \ref{fig3}, the bell state $ \frac{1}{\sqrt2}{\ket{\uparrow}}_1{\ket{\downarrow}}_2+\frac{1}{\sqrt2}{\ket{\downarrow}}_1{\ket{\uparrow}}_2$ is prepared with $\beta_{q1}$, $\beta_{q2}$ evolving according to the intermediate pulse wavepacket $\alpha_{out}(t)$. The corresponding driving frequencies are depicted in Fig. \ref{fig3b}.

Finally, we investigate the effects of coupling strength $\gamma$, $G$, shown in Figs. \ref{fig4a} and \ref{fig4b}, on the fidelity of stimulating a single photon and transferring a local entanglement: $\frac{1}{\sqrt2}\ket{\uparrow}\ket{vac}+\frac{1}{\sqrt2}\ket{\downarrow}\ket{\tilde{\alpha}_{out}}$. In the stimulation process, both $\gamma$ and $G$ have a favorable influence on the fidelity. Minor decoherences serve to amplify this relation. Interestingly though, the coupling strength $\gamma$ in the preparation of the local entanglement has a negative effect on the fidelity. This could be better understood if we account for the superposition characteristic of entanglement. Fig. \ref{fig4c} shows the influence of the qubit dephasing, which is the predominant source of decoherence in a typical three-level quantum interface, on the fidelity $F^{'}$ of tranferring the target entangled state.  

\section{Conclusions}
In this article, we have described a scheme of light-spin interface based on optomechanical resonators and the spin-flexual coupling in a susupended CNTQD. Due to the long lifetime, pure spins can be deterministically prepared and detected. The electrically tunable qubit-phonon coupling strength enables electrical control of the quantum interface between a photon qubit and a single-electron spin qubit. By precisely designing the controlling pulse, specific single-qubit operations as well as two-qubit entanglements locally or nonlocally can be performed. We numerically estimate the state tranfer processes taking into account some decoherence sources. Future optimizations could be the combination designs with such qubits as superconducting charge qubits and Nitrogen-Vacancy (N-V) center spin qubits from other QI schemes.\\

\end{document}